\documentclass[sigconf]{acmart}

\makeatletter
\def\@ACM@checkaffil{
    \if@ACM@instpresent\else
    \ClassWarningNoLine{\@classname}{No institution present for an affiliation}%
    \fi
    \if@ACM@citypresent\else
    \ClassWarningNoLine{\@classname}{No city present for an affiliation}%
    \fi
    \if@ACM@countrypresent\else
        \ClassWarningNoLine{\@classname}{No country present for an affiliation}%
    \fi
}
\makeatother

\AtBeginDocument{%
  }

\usepackage{amsmath,amsfonts}
\usepackage[utf8]{inputenc}
\usepackage{url}
\usepackage{tikz}
\usepackage{textcomp}
\usepackage{graphicx}
\usepackage{gensymb}
\usepackage{dirtytalk} 
\usepackage[euler]{textgreek} 
\usepackage{soul}
\usepackage{comment}
\usepackage{colortbl}
\usepackage{hyperref}

\usepackage[resetlabels,labeled]{multibib}
\newcites{AP}{Analyzed Papers}
 
\newcommand{\se}[1]{{\color{blue}\{SE: #1\}}}
\newcommand{\sv}[1]{{\color{teal}\{SV: #1\}}}
\newcommand{\hidecomments}{
    \renewcommand{\se}[2][orange]{}
    \renewcommand{\sv}[2][orange]{}
}

\hidecomments

\setcopyright{acmcopyright}
\copyrightyear{2023}
\acmYear{2023}
\acmDOI{XXXXXXX.XXXXXXX}

\acmConference[ESEC/FSE 2023]{The 31st ACM Joint European Software Engineering Conference and Symposium on the Foundations of Software Engineering}{3 - 9 December, 2023}{San Francisco, USA}
\acmPrice{15.00}
\acmISBN{978-1-4503-XXXX-X/18/06}

\begin{document}

\title[Pitfalls in Experiments with DNN4SE: An Analysis of the State of the Practice]{Pitfalls in Experiments with DNN4SE: \\ 
An Analysis of the State of the Practice
}


\author{Sira Vegas}
\email{sira.vegas@upm.es}
\orcid{XXX}
\affiliation{%
  \institution{Universidad Politécnica de Madrid}
}

\author{Sebastian Elbaum}
\email{selbaum@virginia.edu}
\orcid{XXX}
\affiliation{%
  \institution{University of Virginia}
}



\begin{abstract}

Software engineering techniques are increasingly relying on deep learning approaches to support many software engineering tasks, from bug triaging to code generation. To assess the efficacy of such techniques researchers typically perform  controlled experiments. Conducting these experiments, however, is particularly challenging given the  complexity of the space of variables involved, from specialized and intricate architectures and algorithms to a large number of training hyper-parameters and choices of evolving datasets, all compounded by how rapidly the machine learning technology is advancing, and the inherent sources of randomness in the training process. In this work we conduct a mapping study, examining 194 experiments with techniques that rely on deep neural networks appearing in 55 papers published in premier software engineering venues to provide a characterization of the state-of-the-practice, pinpointing experiments common trends and pitfalls. Our study reveals that most of the experiments, including those that have received ACM artifact badges, have fundamental limitations that raise doubts about the reliability of their findings. More specifically, we find: 1) weak analyses to determine that there is a true relationship between independent and dependent variables (87\% of the experiments), 2) limited control over the space of DNN relevant variables, which can render a relationship between dependent variables and treatments that may not be causal but rather correlational (100\% of the experiments), and 3) lack of specificity in terms of what are the DNN variables and their values utilized in the experiments (86\% of the experiments) to define the treatments being applied, which makes it unclear whether the techniques designed are the ones being assessed, or how the sources of extraneous variation are controlled. We provide some practical recommendations to address these limitations.

\end{abstract}



\maketitle

\nociteAP*

\section{Introduction}
\label{sec:intro}

The application of deep learning (DL) techniques across the software development life cycle is becoming a thriving  research thread in the software engineering (SE) community. Such emerging techniques, often grouped under labels such as  DL4SE or DNN4SE, have rendered promising results  supporting the automation of activities ranging from  requirements engineering to code maintenance \cite{SE4DLreport}. Similar to other DL application areas, the maturation of frameworks and tools that lowered the bar for the adoption for such technology has facilitated their application in the SE domain. In addition, our community is in an advantageous position in that we can tap into a continuously increasing number of public repositories with various types of software artifacts such as code and tests  that constitute rich data sets on which DL techniques can thrive.
 
To assess such techniques, researchers  perform   experiments in which variables are manipulated in a controlled environment to investigate their impact  over response variables~\cite{juristo2013basics}. Conducting such experiments, however, can be extremely challenging given the number and complexity of variables that may affect a technique that relies on DL. Tens of variables play a fundamental role in how a deep neural network (DNN) is set up as part of an experiment. Some of these variables are inherently complex as they point to  optimization procedures that contain their own set of parameters. Other variables like those associated with  datasets or competing models often point to online resources that  may unsuspectingly evolve. Other variables, like those affecting the  sample used by gradient descent to set the network weights or the proportions of data used for training and testing, are deceptively simple, yet they constitute  sources of randomness that will impact the DNN's performance. Yet other variables that may not be explicitly defined, like the ones defining termination criteria, can have subtle interactions with other variables undermining the implementation of the intended experimental constructs. The key takeaway is that  when evaluating the application of a DL technique to a problem through an experiment, the lack of careful consideration of a complex set of variables can  dramatically impact the findings. 

The \textbf{goal} of this paper is to begin understanding the extent to which experiments on DNN4SE techniques are addressing the distinct experimental challenges introduced by DNNs.  

In pursuing that goal we make four contributions.  

\vspace{0.03in}\noindent
I) We contribute a characterization and analysis of the state-of-the-practice of experimentation with DNN4SE by addressing a fundamental question: \textbf{RQ1: To what extent are DNN4SE  experiments specified in papers?} To answer that question, in Section~\ref{sec:analysis_papers}, we present a systematic mapping study~\cite{kitchenham2023segress} of 55 papers from ICSE, FSE, and TSE from 2018-2021 that apply DL techniques to automate SE tasks. Building on a cause-effect model of the experimental space and the variables relevant to DNNs, we determine the degree to which the variable space in the experiments was specified by each paper. We find that while most experiments clearly identify, for example, the response variables (76\%) and training data (69\%), none describe their complete space of variables. Furthermore, most experiments lack in critical  aspects like the choice for experimental design  to control the sources of variability (30\%) and the use of even descriptive statistics as part of the results analysis and interpretation (56\%). This lack of specificity is not just an under-reporting issue, but it  reflects a limited consideration of fundamental experimental aspects that threaten the validity of the findings.

\vspace{0.03in}\noindent
II) Given the community ongoing efforts for sharing artifacts~\cite{artifactsInSE}, we extend the previous characterization through \textbf{RQ2. Do shared artifacts improve the specifications of DNN4SE  experiments provided in the papers?} Section~\ref{sec:analysis_artifacts} contributes an analysis of the artifacts associated with the subset of papers that earned ACM artifact badges, increasing the depth of  analysis to include code, data, and documentation.  As expected,  artifacts complement some but not all aspects presented in the papers, especially  the definition of variables and the training and test data, all of which are necessary to operationalize the experiments. However, we also find that 68\% of the experiments reported in the artifacts present inconsistencies when compared with the corresponding paper, ranging from the loss function  to the testing data being used.  This is problematic because the additional effort invested   to prepare artifacts  to further support the experiments often raises doubts about which portions of the papers and the artifacts  are to be trusted. 

\vspace{0.03in}\noindent
III) We contribute an analysis of why these findings matter through \textbf{RQ3. What are the implications of the previous findings about the under-specification of DNN4SE experiments?}  Section~\ref{sec:issues} summarizes these implications. First, by failing to clearly define factors and treatments in 86\% of the experiments, it is unclear whether most experimental results are caused by the intended constructs or by other variables that were not operationalized correctly. In the best of cases, one could argue that those unspecified variables in the papers are controlled when the experiments are performed. However, our analysis of artifacts reveals that that is rarely the case. Second, even when variables are specified it is often unclear how they are controlled to  establish causality. We find that 62\% of experiments  account for sources of randomness related to the dataset, and none  controlled for other sources of training randomness  by, for example, performing multiple training runs or varying the DNN initial weights. Third, we find that 56\% of experiments identify relationships between independent and dependent variables based on single observations which is suspect as it ignores any experimental fluctuation.
 
 \vspace{0.03in}\noindent 
IV) \textbf{Recommendations.} We are not the first community challenged by the DNN complexity. The AI community has developed various checklists to mitigate common ML experimental pitfalls~\cite{AAAI2023checklist,ML2020checklist,NeurIPS2022checklist}.
Similarly, the SE community has developed a  body of knowledge to assess and improve the quality of the experiments we conduct (see related work in Section~\ref{sec:related_work}). However, as it shall become clear from our RQ1-RQ2-RQ3 findings, there is a distinct and urgent need for the SE community to become much more cognizant of how to manage the space of variables particular to the DNN domain. Towards that end, we recommend actionable practices to manage the challenges in DNN4SE experimentation (Section~\ref{sec:recommendations}) that, if adopted, can alleviate many of the concerns   we encountered. For example, simply  conducting multiple DNN training runs to control for randomness could benefit almost all experiments, performing more meaningful comparisons over multiple observations to account for experimental variability in DNNs  could benefit from 56\% to 87\% of the experiments, and  standardizing a minimal specification of the space of DNN training variables and providing partial automation for synchronizing the paper and artifact content of DNN4SE experiments could benefit 96\% of the experiments. 
\section{DNNs' Experimental Variables }
\label{sec:DNNs}

\begin{figure}[bt]
      \centering
      \includegraphics[width=\linewidth]{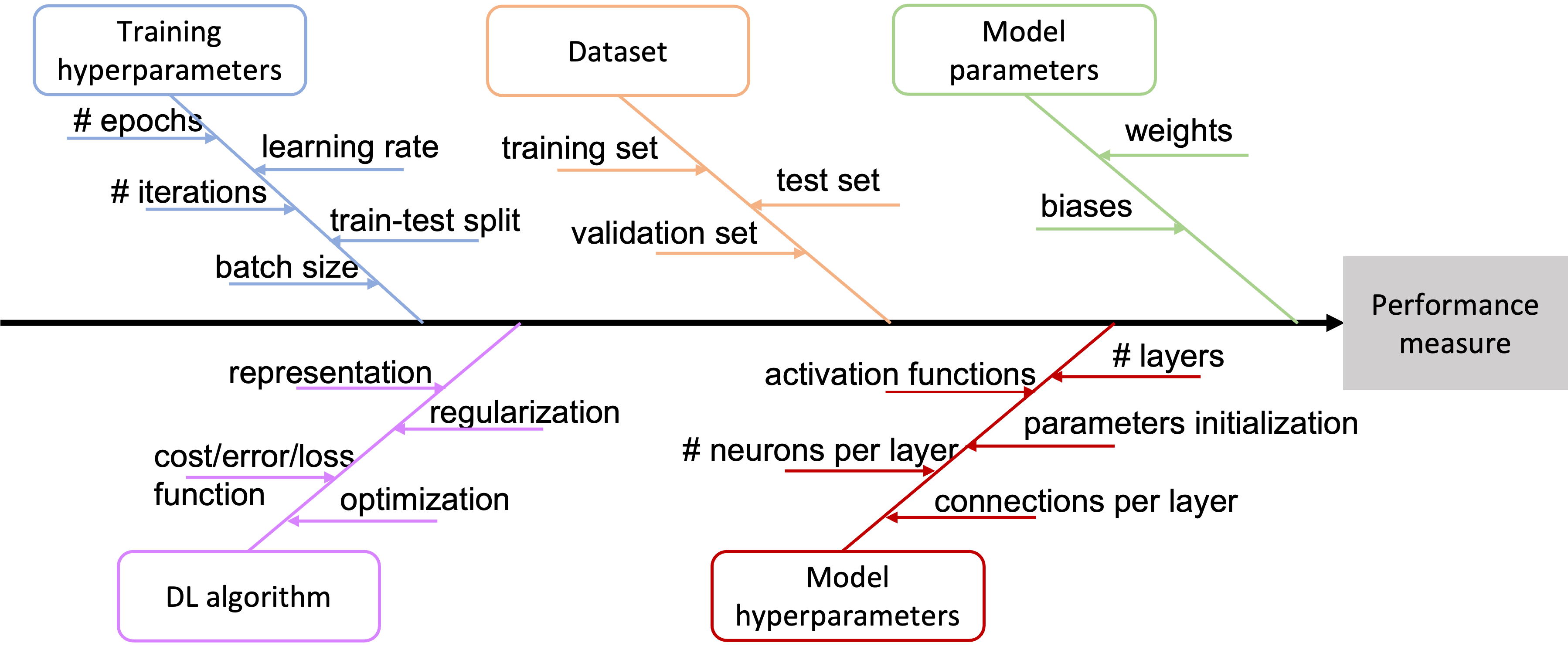}
      \caption{A cause-and-effect diagram for   5 groups of variables   in experiments with DNNs.}
      \label{fig:fishbone} 	
    \end{figure}

Machine learning (ML) is a subfield of Artificial Intelligence  that aims to enable computers to learn from experience~\cite{goodfellow2016deep,mitchell2019machine}. ML algorithms build a model based on sample (training) data  to make predictions  without being explicitly programmed to do so~\cite{samuel1959studies}. DL is a type of ML technique supported by neural networks that have a deep architecture as per their constituting layers ~\cite{goodfellow2016deep}. 
The training of these DNNs consists of adjusting its \textbf{model parameters}, using a \textbf{deep learning  algorithm} controlled by a set of \textbf{training hyperparameters} and \textbf{model hyperparameters}, using a \textbf{dataset}~\cite{goodfellow2016deep}.   
We will later use these 5 groups of variables associated with DNNs,  illustrated   through a cause-effect diagram in Figure~\ref{fig:fishbone},  as a basis for the analysis of experiments. 

The DNN  overall architecture is defined by the \textbf{model hyperparameters} and includes 5 variables. DNNs consist of interconnected \textit{neurons} grouped in \textit{layers}. There is always an input layer  that accepts inputs  and an output layer that provides the output, and at least two  hidden layers between them. There are different \textit{layer types}, and how those layers are connected define the higher-level architecture of the DNN (e.g., feed-forward, CNN, RNN, LSTM). Every connection between 2 neurons has a weight which regulates how much of the initial value will be forwarded to a given neuron. Each neuron has an associated value, called the bias. Weights and biases need to be \textit{initialized} prior to training. The sum of the products of the inputs and respective weights, plus the bias, are then provided to an \textit{activation function} to produce a neuron's output. A forward pass is the set of calculations that take place when the input travels through the DNN to the output.   

The neuron \textit{weights} and \textit{biases} constitute the two variables defining the DNNs \textbf{model parameters}. They are initialized before training and reset during the subsequent phases of the training process.  
    
A \textbf{dataset} is a collection of inputs and outputs. At least two types of datasets are required: \textit{training} and \textit{test}. The training set is used to adjust the model parameters during training. The test set is used to check how well the algorithm performs on data that has not seen before, and  it is intended to estimate the generalization error. The training set can be further divided into a training and a \textit{validation set}. This validation set can be used to get an estimate of model skill while tuning its hyperparameters.  
 
The \textbf{DL algorithm} is defined through 4 variables~\cite{goodfellow2016deep}: a \textit{representation} for encoding the elements in the dataset, a \textit{function measuring the error} between the value predicted by the model and the real value,  an \textit{optimization} procedure to minimize the training error (e.g. stochastic gradient descent, Nesterov momentum, Adam), and \textit{regulatization} strategies to reduce the generalization (test) error (e.g. dropout, data augmentation, early stopping). 

During training, the DL algorithm's behaviour is controlled through 5 \textbf{training hyperparameters}~\cite{goodfellow2016deep}. 
The \textit{batch size} defines the number of training samples to consider per training \textit{iteration}. Depending on the batch size, multiple iterations will be needed to go through the entire training set. The \textit{number of epochs} defines how many times the algorithm will go through a dataset. The \textit{train-test split} defines on what portion of the data training is performed. Given a batch, the network performance (measured as a function of error/cost/loss) is used to drive the backpropagation (the reverse of a forward pass using gradient descent) to update the network  weights and biases to minimize this error. The \textit{learning rate} specifies how much to update the model in response to the estimated error.

Albeit simplified and limited for exposition, this section highlights the vast space of  variables involved in training a DL system, where each  one can take an increasing number of values. These variables also have many subtle interdependencies (e.g., the batch and epoch size often depend on the parameter initialization, the loss function depends on the architecture, the architecture depends on the data dimensionality). Confounded with the multiple sources of randomness involved in the DL training process (e.g., different train-test partitions, different sample batches   being selected, different portions of the network being targeted for regularization, different supported hardware), defining and conducting robust experiments is intrinsically challenging.


\section{Analysis of Papers}
\label{sec:analysis_papers}
 
In this section we answer \textbf{RQ1} by providing an overview of the state of the practice in performing experiments where DNNs are utilized to address SE challenges (DNN4SE). We characterize the growing number of experiments being carried out in this domain and identify some overarching  limitations across those experiments.

\subsection{Scope of Analysis}
\label{sec:scope}
 
\begin{figure}[bt]
      \centering
      \includegraphics[width=\linewidth]{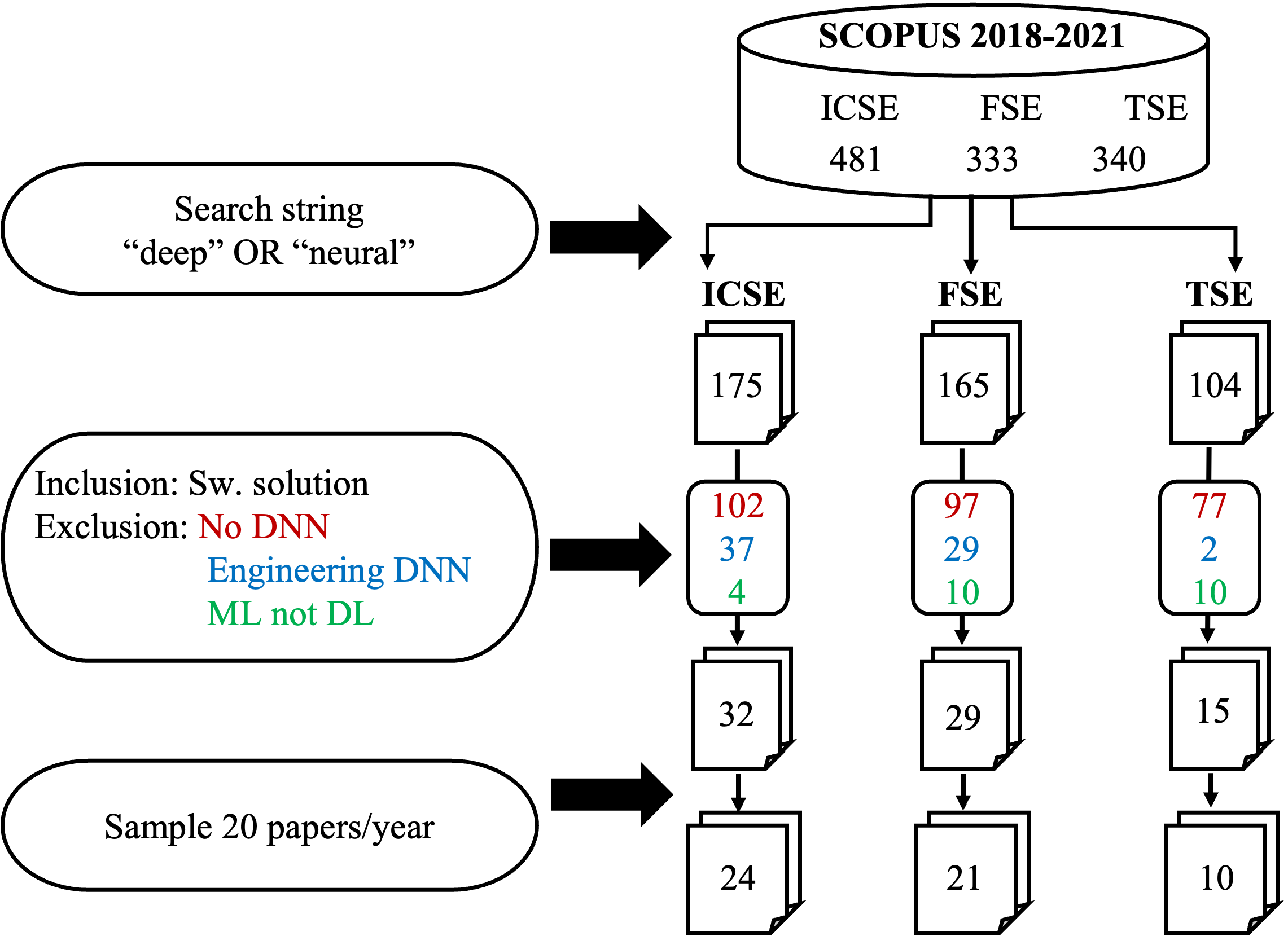}
      \caption{Paper search and selection process. }
      \label{fig:search} 	
    \end{figure}

We have performed a semi-automated search of papers reporting experiments with DNNs developed to solve SE tasks. Figure~\ref{fig:search} summarizes the search and selection process. We have shared in the paper repository the outputs of each step.

In a first automated step, during January 2022, we searched SCOPUS\textsuperscript{TM} using the string \say{deep OR neural} in all fields. The search was limited to full papers from the technical track of the International Conference on Software Engineering (ICSE) and the Joint European Software Engineering Conference and Symposium on the Foundations of Software Engineering (ESEC/FSE), and papers published in IEEE Transactions on Software Engineering (TSE), covering the period 2018-2021.   We decided to favor the flagship conferences ICSE and FSE because we believe they include the latest work in DL and appear at the top of various ranks\footnote{https://csrankings.org, http://portal.core.edu.au/conf-ranks}. Similarly, we selected TSE because it has the highest impact factor among SE journals\footnote{ https://www.scimagojr.com, https://jcr.clarivate.com/jcr/home}.  The search resulted in 444 out of 1154 published papers.

Next, we examined the papers to exclude the ones that did not cover techniques using DNNs to address software engineering challenges. The examination was conducted by one of the senior researchers authoring this paper with expertise in empirical software engineering and DNN development.  This process led to the exclusion of 276 papers that did not include a DNN (e.g. a DNN-solution is part of the related work in a paper), 68 papers that focus on improving the engineering of DNN-solutions (e.g. testing of DNN-solutions), and 24 papers that use ML mechanisms but   not DNNs (e.g., shallow networks). When papers that did not clearly fit in existing categories were found, they were examined and discussed jointly by both authors. The remaining 76 papers report experiments with DL-based software to address SE challenges. 
 
Table~\ref{tab:experiments_found} shows the paper count distribution over the years and venues. We can see that the number of papers in this area is steadily increasing over the last few years, from 6 papers in 2018 to 41 papers in 2021. For the subsequent analysis we selected every paper identified as within scope from 2018 to 2020, and given the larger number of relevant papers published in 2021 (from 16 in 2020 to 41  in 2021), we randomly sampled 20 papers from 2021. This sampling was necessary to control the cost of the study given that just data extraction time per paper was approximately 4 hours per person (we later describe the analysis costs per paper and artifact). This gave us  a total  of 55 papers spanning four years to analyze.  

\begin{table}[bt]
  \centering
  \caption{Number of papers within scope analyzed / published, and experiments analyzed (in parenthesis)}
  \label{tab:experiments_found}
  \begin{tabular}{lccccc}
  \toprule
	&	\textbf{ICSE}	&	\textbf{ESEC/FSE}	& \textbf{TSE}	 & \textbf{Total} \\ 
	\midrule
	
    \textbf{2018}	& 2/2 (5)		&	4/4 (11)	&	 0 (0) & 6/6 (16) \\
    \textbf{2019}	& 8/8 (27)	&	4/4 (9)	& 1/1 (6) & 13/13 (42) \\
    \textbf{2020}	& 7/7 (25)	&	6/6 (21)	& 3/3 (14) & 16/16 (60) \\
    \textbf{2021}   & 7/15 (21)	&    7/15 (32)	&  6/11 (23)  &  20/41 (76) \\ 

    \midrule
    
    \textbf{Total}   & 24/32 (78)    &  21/29 (73)     & 10/15 (43) & 55/76 (194) \\

\bottomrule
\end{tabular}
\end{table}

\subsection{Analysis Process}
\label{sec:analysis_process}

To have a consistent data extraction process from the papers, we defined a set of scoping and analysis guidelines.

First, for each paper, we initially considered just the contents of the published paper. At this early examination stage we did not peak into  artifacts that may be associated with the paper like code repositories as we wanted to have a common baseline of materials among all covered papers. This also made the analysis cost more viable at the first stage of the study. In addition, for each experiment  in a paper we bounded the analysis to the DNN portions. That is, when we found  experiments  comparing the performance of DNNs against other type of approaches that employ traditional SE approaches or humans, we deemed those portions of the experiment as already understood by the community and only focused on the portions including DNNs. 

Second, we controlled for three common sources of uncertainty we faced when analyzing the papers. To control for different reporting styles, we examined the papers in their totality as we often found portions of the experiments distributed and modified throughout the paper. For example, we found instances where the experimental designs are  sprinkled through background, approach, study design, and results. To control for DNN usage types, we only considered papers that use DNNs to perform either complex data encodings or function as a model. Third, to control for different levels of detail across experiments, we decided to account for all experiments mentioned in the paper, even if marginally reported. 

Given the previous guidelines,  the analysis process started with both authors jointly developing an initial characterization schema for the experiments. This   schema is based on the steps of the experimental process ~\cite{gomez2014understanding,montgomery2019design,wohlin2012experimentation}, although we   adapted those steps to account for the types of variables found in DNNs such as the model hyperparameters, the training hyperparameters, the DL algorithm, the dataset(s), and the model parameters (as per  Figure \ref{fig:fishbone}). Then, both authors conducted a refinement and calibration cycle by extracting the information from all the experiments reported in the 17  ICSE papers from 2018 to 2020 according to the schema. This resulted in a refined schema and a more consistent evaluation process. Finally, each remaining paper was examined by just one author. However, when experiments did not fit the schema, introduced new DNN elements, or had ambiguous specifications, they were examined and discussed jointly by the researchers. There were 8 of such joint examinations, lasting between 1-3 hours, which often triggered the re-examination of previously evaluated papers to ensure their consistent analysis. 

Table~\ref{tab:example} exemplifies the analysis we performed for an experiment. The goal of this paper~\citeAP{paper39} is to perform log-based anomaly detection. The proposed DNN receives as input a sequence of log events and predicts whether the sequence is an anomaly. The experiment evaluates the performance of the proposed DNN in terms of precision, recall, and F1, comparing it against 4 other approaches (none are DNNs). Column 1 and 2 show the steps and aspects of the experimental process, and column 3 assesses to what extent the information has been  identified (Fully, Partially, or Missing). The last column provides an explanation of what is lacking. For this experiment, we have been able to find all information related to research hypotheses, DL algorithm, response variables and test set characteristics. For this reason, their final assessment is \say{Fully} addressed. We have not been able to find any information related to model parameters nor statistics, and therefore, their final assessment is \say{Missing}. For the rest we have not able to find some information, therefore, the final assessment is \say{Partially} addressed. A detailed description of the classification criteria and its application to all the 55 analyzed papers is available in the  repository (Section~\ref{sec:repo}).

\begin{table*}[bt]
\caption{Assessing of a sampled experiment~\cite{paper39} specification in terms of Fully addressed, Partially addressed, or Missing.}
  \label{tab:example}
  \small
  \begin{tabular}{llll}
    \toprule
    \textbf{Step} & \textbf{Aspect} &  \textbf{Assessment}  & \textbf{What is lacking}\\ 
    \midrule
S1. Hypotheses formulation & Research hypotheses	& Fully   &   	 \\	
S2. Variables identification & 	Model hyperparameters &	 Partially  & Missing hyperparameters for initialization \\
 & 	Model parameters &	 Missing  & Missing a pointer   to where they can be found \\
& 	DL algorithm &	Fully   &    \\
& 	Training hyperparameters & Partially  & Missing train-test split and learning rate	   \\
& 	Training data &	Partially   & No information about a dataset for confidentiality reasons\\
S3. Operationalization & Factors and treatments  & Partially    & Some model and training hyperparameters are missing, not all  \\
& & & training data available, and model parameters are missing	 \\
& Response variables   &    Fully   &   \\
S4. Design & Choice of design	&	Partially  	& No analysis of sources of randomness, whether they have been   \\
& & & controlled, and if so, the mechanism used \\
& Instrumentation	&  Partially  & One test set is missing due to confidentiality issues. Software \\
& & & environment is not defined. Measuring instruments and \\
& & &  procedure can be deduced but are not defined \\
S5. Objects selection & Test set chars.	& Fully  &    \\
S6. Analysis \& interpretation & Descriptive statistics	  & Missing  & No descriptive statistics reported \\
& Inferential statistics	&  Missing & No inferential statistics reported\\
S7. Validity evaluation	& Validity threats  &  Partially   & Missing internal, construct and conclusion \\
  \bottomrule
\end{tabular}
\end{table*}

\subsection{Findings}
\label{sec:findings_papers}

Table~\ref{tab:state_practice} summarizes the findings for the 194 experiments analyzed across the 55 identified target papers. It is encouraging to find that most experiments  specify at least to some extent the response variables, the research hypotheses, and the training and test set data.  However, the rest of the experimental aspects tend to be under-specified. We find that 50\% (7 out of 14) of the aspects are  partially addressed, while another 21\% (3 out of 14) of the aspects are missing among the experiments detailed in the papers. 

\begin{table}[bt]
  \setlength{\tabcolsep}{3pt}
  \centering
  \caption{Characterization of 194 experiments with DNNs}
  \label{tab:state_practice}
   \small
  \begin{tabular}{llccc}
    \toprule
    \textbf{Step} & \textbf{Aspect} &  \textbf{Full} &  \textbf{Partial} &  \textbf{Missing} \\  
    \midrule

S1	&	Research hypotheses	&	\textbf{76\%}	&	0\%	&	24\%	\\				
S2	&	Model hyperparam.	&	7\%	&	\textbf{85\%}	&	8\%	\\				
	&	Model parameters	&	2\%	&	0\%	&	\textbf{98\%}	\\				
	&	DL algorithm	&	26\%	&	\textbf{72\%}	&	2\%	\\				
	&	Training hyperparam.	&	19\%	&	\textbf{73\%}	&	8\%	\\				
	&	Training data	&	\textbf{69\%}	&	27\%	&	4\%	\\				
S3	&	Factors and treatments	&	14\%	&	\textbf{82\%}	&	4\%	\\				
	&	Response variables	&	\textbf{76\%}	&	18\%	&	6\%	\\				
S4	&	Choice of design	&	0\%	&	\textbf{70\%}	&	30\%	\\				
	&	Instrumentation	&	2\%	&	\textbf{97\%}	&	1\%	\\				
S5 	&	Test set characteristics	&	\textbf{59\%}	&	19\%	&	22\%	\\				
S6	&	Descriptive statistics	&	10\%	&	34\%	&	\textbf{56\%}	\\				
	&	Inferential statistics	&	12\%	&	1\%	&	\textbf{87\%}	\\				
S7	&	Validity threats 	&	2\%	&	\textbf{79\%} &	19\%	\\		

  \bottomrule
\end{tabular}
\end{table}

We find that essential aspects are missing in  most experiments. For example, for the model parameters  to be fully addressed, we required a pointer to a repository where they could be found. Such pointer was lacking for 98\% of the experiments. For the choice of (experimental) design   to be fully addressed we required a description of what variables are manipulated or controlled and how, yet 30\% of the experiments did not have it. For the analysis and interpretation (S6)  to be fully addressed we required descriptive and inferential statistics, yet they were missing for 56\% and 87\% of the experiments respectively. These results at least raise doubts about whether most of the papers are: 1) implementing the construct they are intending, 2) performing meaningful assessments given the experimental noise that is not accounted for by the analysis and interpretation, and 3) establishing causality given the limited amount of control over the large and complex space of variables to be specified.


\section{Analysis of Artifacts}
\label{sec:analysis_artifacts}

In Section~\ref{sec:analysis_papers} our analysis of papers revealed that the under-specification of  experiments with approaches that use DL to address SE problems is pervasive. Still, given our community growing practice towards artifact sharing~\cite{artifactsInSE} and the nature of DL experiments (i.e., large open datasets, common architectures, standard APIs), it seems reasonable to ask whether the missing portions of the experiments specifications appear in the shared artifacts. This is also important as it may let us understand if the problem is just one associated with how experiments are reported or if there is a deeper concern about how the experiments are being conducted. 

We begin to answer \textbf{RQ2} through an analysis of the artifacts associated with those papers to assess the degree to which the under-specification in the papers is complemented by the associated artifacts, and whether the  design and analysis limitations identified are mitigated by the artifacts.

\subsection{Scope of  Analysis}
\label{sec:scope_artifacts}

Forty-eight out of 55 papers point to some kind of external artifact. 
A cursory analysis of those artifacts reveals that their content (from just readmes plus code to experimental results and even new experiments), availability (from broken links to pointers to   private repositories or Zenodo), and quality (from a model dump without any explanation to those including a code base to reproduce the results in the paper) had too much variance to define a standardized analysis that would render meaningful findings. This finding is consistent with recent reports on artifact quality~\cite{artifactsInSE}.

Thus, to get a more precise estimate of the degree of under-specification when considering artifacts, we reduce the scope of analysis to the artifacts associated with the 9 papers (including a total of 44 experiments) that earned at least one of the ACM artifact badges\footnote{ACM defines three  badges: Artifacts Evaluated (successfully completed an independent audit, with two levels: Functional and Reusable), Artifacts Available (available for retrieval), and Results Validated (results obtained by a team other than the original, with two levels: Results Reproduced and Results Replicated). The 9 papers we analyzed earned the Artifacts Available badge, and three of  them also earned  the Artifacts Evaluated (two Reusable~\citeAP{paper22,stateformer} and one Reusable and Functional~\citeAP{paper82}).} \cite{ACMbadges}.
This reduced scope allows us to focus more deeply on papers vetted by a conference committee according to established guidelines regarding their completeness and quality.

\subsection{Analysis Process}
\label{sec:analysis_process_artifacts}

We analyzed all artifacts with the following process. First, we examined the readme files and other introductory documentation to get a broad sense of what the artifact was meant to provide. Second, we systematically explored the artifact  directories and their contents to identify the resources of information to collect the data required for Table~\ref{tab:example}. Third, we analyzed the code broadly construed to include Python or C code, configuration files, and batch scripts. The analysis was first meant to map each experiment reported in the paper to the items in the artifact. Although conceptually simple, this analysis process was nothing but straight-forward as the artifact structure rarely matched that of the paper (where the experiments are reported). In most cases, we had to recover  portions of one or multiple experiments from undocumented code. This required multiple inspections of the code, running portions of it to confirm that what was learned from the static inspections, and referencing back the findings to the information in the paper.  Fourth, for each experiment identified in the artifact, we collected metadata such as the one reported in Table~\ref{tab:example} (more details about the information collected are provided in the repository described in Section~\ref{sec:repo}). During this step we also determined whether the artifact improved or complemented the information provided in the paper, and recorded any inconsistencies we found between them. These steps required approximately 8 hours per  paper (\citeAP{stateformer} was an exception given the number of experiments reported). The difficulties in this process, particularly in the third and fourth steps, and the time allocated per paper, forced us to be conservative in our assessment, only judging an artifact experiment to be  incomplete or inconsistent with the paper when we had a high certainty that that was the case. Still, these sources of uncertainty in our analysis constitute a threat to the validity of our findings (Section~\ref{sec:validity}) that we mitigate by sharing our data (Section~\ref{sec:repo}).

\subsection{Findings}
\label{sec:findings_artifacts}
 
Table~\ref{tab:artifacts} summarizes our findings for the 44 experiments from papers that earned ACM badges. 
The columns under Improvements contain the \% of experiments exhibiting gains across the specification levels (i.e., $PA\rightarrow FA$ means improvement from partially addressed to fully addressed), while the columns under Constant show the aspects of the experiments that remained unchanged.

\newcommand{\myarrow}[1]{%
\parbox{#1}{\tikz{\draw[->](0,0)--(#1,0);}}
}
 
\begin{table}[bt]
{ 
  \caption{Characterization of (44) experiments that earned ACM Artifact Badges. } 
  \label{tab:artifacts}
    \small
  \setlength{\tabcolsep}{2pt}
  \begin{tabular}{ll|cccc|ccc}
    \toprule
    & & \multicolumn{4}{c|}{Improvements} & \multicolumn{3}{c}{Constant}\\
Step	&	Aspect	&	PA$\myarrow{.08cm}$FA		&	M$\myarrow{.08cm}$PA		&	M$\myarrow{.08cm}$FA		&	PA$\myarrow{.08cm}$PA		&	M		&	PA		&	FA		\\	\midrule
S1	&	Research hypotheses	&	0\%	&	0\%	&	0\%	&	0\%	&	18\%	&	0\%	&	82\%	\\	
S2	&	Model hyperparam.	&	68\%	&	0\%	&	7\%	&	0\%	&	0\%	&	5\%	&	20\%	\\	
	&	Model parameters	&	0\%	&	0\%	&	9\%	&	0\%	&	82\%	&	0\%	&	9\%	\\	
	&	DL algorithm	&	39\%	&	0\%	&	7\%	&	0\%	&	0\%	&	13\%	&	41\%	\\	
	&	Training hyperparam.	&	59\%	&	0\%	&	7\%	&	0\%	&	0\%	&	14\%	&	20\%	\\	
	&	Training data	&	7\%	&	0\%	&	7\%	&	0\%	&	2\%	&	4\%	&	80\%	\\	
S3	&	Factors \& treatments	&	0\%	&	2\%	&	5\%	&	52\%	&	0\%	&	30\%	&	11\%	\\	
	&	Response variables	&	2\%	&	3\%	&	9\%	&	0\%	&	2\%	&	0\%	&	84\%	\\	
S4	&	Choice of design	&	0\%	&	0\%	&	0\%	&	0\%	&	39\%	&	59\%	&	2\%	\\	
	&	Instrumentation	&	2\%	&	7\%	&	0\%	&	5\%	&	2\%	&	75\%	&	9\%	\\	
S5	&	Test set chars.	&	0\%	&	0\%	&	7\%	&	0\%	&	43\%	&	30\%	&	20\%	\\	
S6	&	Descriptive statistics	&	0\%	&	2\%	&	0\%	&	0\%	&	64\%	&	23\%	&	11\%	\\	
	&	Inferential statistics	&	0\%	&	0\%	&	0\%	&	0\%	&	95\%	&	0\%	&	5\%	\\	
S7	&	Validity threats	&	0\%	&	0\%	&	0\%	&	0\%	&	27\%	&	71\%	&	2\%	\\	

    \bottomrule
    \end{tabular}
} 
\end{table}

Overall, and as expected, we find that considering the artifact   consistently improves the specification of some portions of the  experiments but not others. The improvement is particularly noticeable in the variable identification step (S2) where many experiments that were Partially Addressed (PA) become Fully Addressed (FA). More specifically, the DL algorithm, model and training hyperparameters and the training data become fully addressed in 87\%, 95\%, 86\% and 94\% of the experiments, respectively\footnote{The exceptions are two optimization experiments missing from  the artifact's code (E4~\citeAP{paper30} and E1~\citeAP{lightweight}), and 4 experiments in a paper that are missing the training code~\citeAP{paper22}.}. The model parameters (also part of S2) show a modest 9\% gain caused by the artifact for just one of the papers (\citeAP{paper22}). Under operationalization (S3), the response variables also improve, becoming full for 95\% of the experiments, while  factors and treatments show some improvement for 59\% of the  experiments but still remains partially addressed for 84\% of the experiments. These operationalization improvements were also expected as the code  must assign values to the independent variables and measure the dependent variables to assess the experimental outcome. The rest of the aspects, which are more closely associated with the experimental design  and analysis than the implementation, showed slight or no improvement. The instrumentation showed an improvement for 14\% of the experiments, test set characteristics for 7\%, descriptive statistics for 2\%, and research hypotheses, choice of design, inferential statistics, and validity threats showed no improvement. In summary,  considering the artifacts improved the aspects associated with S2, but the rest of weak spots identified in the papers remain.

Our inspection also reveled  several incomplete artifacts. We found that papers pointing to a piece of information that is not accessible in the artifact, either because it is missing from the artifact (e.g., paper~\citeAP{paper26} mentions that the artifact includes ``all model information'', but the model parameters are missing) or because it requires special permissions or has   broken links  (e.g., paper~\citeAP{stateformer} contains dropbox links to training data that need permission). 

More problematic, however, the inspection of the artifacts revealed many cases where \textit{the experiments in the artifact and the experiments reported in the paper are inconsistent}. We found that most artifacts contained pieces of code representing variations of the experiments reported in the paper. This in itself is not a major source of concern as one may conjecture that these variations corresponded to different configurations  explored during the investigation and development of the proposed techniques, configurations that perhaps were not properly labeled or cleaned from the shared code base. What is concerning, however, are the cases where the artifact does not have a single experiment variant that matches the  experiment reported in the paper. 

When comparing papers and artifact content, we find that 78\% of the papers and 68\% of the experiments show inconsistencies. For example, \citeAP{paper34} mentions that the loss function used is binary cross-entropy, while the sigmoidal cross-entropy function is used in the artifact code. Paper~\citeAP{paper82} mentions the programs used as test sets for the paper, but the artifact contains a different set of programs. Paper \cite{paper62} makes a reference to grid search, which is absent in the artifact. Paper~\citeAP{syntaxguided} mentions that the Adam optimizer is used, but the code also contains AdamW. Again, our analysis was conservative and the time dedicated to explore the artifacts was bounded, so it is reasonable to expect the inconsistencies found are likely an underestimate of the ones present. We also found artifacts that were at times inconsistent with themselves. For example, \citeAP{stateformer} provides generous supplementary information in the form of an online appendix that contains information related to experiments that are not reported in the paper, but these show the same inconsistencies with the code that the paper has regarding model hyperparameters and training data. Similarly, \citeAP{lightweight} does not mention in the paper the number of epochs used, and there are two values for it in the configuration file contained in the artifact. 

It is important to emphasize that the analysis of the artifacts provides further evidence that the limitations we have identified in these experiments go beyond under-reporting  problems. The lack of specificity in fundamental experimental design and implementation details reflect deficiencies that can have severe implications for the findings. We delve into these implications next.


\section{Implications}
\label{sec:issues}

The previous sections characterized the degree of under-specification in DL experiments to address SE problems when considering papers and artifacts. We found that the most affected experimental aspects are the analysis and interpretation of results, the design and the operationalization of factors and treatments. In this section we answer \textbf{RQ3} by deriving the implications of under-specifying those aspects from the perspective of conclusion, internal,  and construct validity of the experimental findings~\cite{shadish2002experimental}.

\subsection{Is there a Relationship between the Response Variable and the Factor(s)? (Conclusion Validity)}
\label{sec:conclusion}

The experiments analyzed include 3 types of analysis to determine if there is a relation between the factors and the response variable.

We have found that 56\% of the reviewed experiments resort to \textbf{comparing single data points}. This is problematic because it assumes that a single observation on the effect of the treatment will be a good estimate of the mean effect of that treatment, basically ignoring fluctuations due to experimental errors (this will be further discussed in Section~\ref{sec:internal}). For example, \citeAP{paper09} proposes a DNN that given as input a set of \textit{may} links between communicating objects in two Android applications, outputs the probability that such  links exists. The proposed approach is compared against 3 simpler DNN architectures as baselines. Based on the comparison of the values obtained from the test set for the four alternatives, the paper concludes that the best option is their proposed model (the most complex one), with  response variables values of  0.931 (F1), 0.991 (AUC) and 0.992 (Kruskal's \textgamma). However, the results of the second best option are 0.920, 0.988 and 0.989 respectively. Note that a mere standard deviation of 0.0155, 0.0045 and 0.0045 in the response variables (assuming a sample size of 30) will invalidate the conclusion. Note that often the single point comparison is the result of a problem with the design of this experiment, which does not control, for example, for random sources of variation that would have required multiple runs and hence resulted in multiple values to perform a statistical comparison that accounts for variability. A variant of this problem is manifested in~\citeAP{automaticfeature}, which proposes a DNN that receives a code function as input and predicts whether it is vulnerable. The paper computes the performance of three techniques over multiple Android applications in terms of precision, recall, F-measure, and AUC. However, it then resorts to  count the number of projects in which each technique has shown better results and  compares those single values losing an opportunity to perform a more meaningful comparison. 

We find that 31\% of the experiments perform a \textbf{comparison of means}. This is  stronger than using single data points, but still insufficient to guarantee that the differences found in the sample   can be extrapolated to the population the sample represents.
For example, \citeAP{paper59} proposes a DNN that takes as input color pictures of source code files to predict whether they contain a fault. It uses 10 test sets corresponding to open source projects to assess the proposed approach against 4 existing techniques as per their mean F-measure for the different projects, and concludes that  \say{the proposed DTL-DP shows significant improvements on the state-of-the-art in cross-project defect prediction}. Yet, there is no analysis that considers the variability observed on the collected  measures, even though the F1 values showed large variability. 
To better understand the implications of this oversight we perform a statistical analysis with the data reported in Table 4 of the paper. Let's assume that the statistical null hypothesis (H0) is: ``There is no difference in F-measure between the different approaches examined'', and that the  design is a 1-factor 5-levels experiment (inferred from the  design description). The 1-way repeated measures ANOVA shows that we can reject the null hypothesis (p$<$0.01). The follow-up Bonferroni multiple comparisons test shows that the proposed approach has a better performance than three of the competing ones, but  similar to one of them (DBN-CP, Cohen's d=0.3). This example illustrates that relationships identified through means may not necessarily be generalizable to the population.

Only 13\% of the papers we reviewed identify the potential relationship through \textbf{inferential statistics}, meaning that the obtained results can be generalized from the experiment sample to the population it represents. For example, \citeAP{whichvariables} proposes a DNN that given as input a code snippet that needs to be logged, suggests which variables should be logged. Their proposed approach is compared against 5 baselines for 9 different projects in terms of accuracy, mean reciprocal rank and mean average precision. Data is analyzed with a Wilcoxon signed-rank test (considering the 9 scores, one per project), and Cliff's Delta effect size is computed. In all cases, the improvement of the proposed approach is statistically significant, with a large effect size.  

\subsection{Is the Relationship Causal? (Internal Validity)}
\label{sec:internal}

In an experiment, the extent to which extraneous variables are accounted for in the design will define the strength of the causality link~\cite{juristo2013basics}. Table~\ref{tab:dealing_extraneous} shows some of the established recommended mechanisms to deal with extraneous variables~\cite{altman2015sources,juristo2013basics,montgomery2019design,wohlin2012experimentation}, which depend on the nature of the extraneous variable being controlled. For example, when the variable is known, measurable, and controllable, then we can address it either holding it constant or by incorporating it an experimental factor (e.g. dataset); and when the variable is known and measurable but not controllable we can use blocking to control its impact (e.g. random training/test split). Such mechanisms naturally apply to DL.

\begin{table}[bt]
  \caption{Extraneous variables and how to deal with them}
  \label{tab:dealing_extraneous}
  \setlength\tabcolsep{1.5pt} 
  \small
  \begin{tabular}{cllll}
  \toprule
	& \multicolumn{3}{c}{\textbf{Characteristics}} & \textbf{Mechanism}  \\ \cmidrule{2-4}
	
	\textbf{Case} & Known & Measurable & Controllable & \\
	\midrule
	I & No & - & - &  Randomization \\
	II & Yes & No & - & Case I + Replication \\
	III & Yes & Yes & No & Case II + Statistical adjustment \\
	IV & Yes & Yes & Partially & Case III + Blocking \\
	V & Yes & Yes & Yes & Case IV + Held-constant \\
	& & & &  Incorporate as factor \\
\bottomrule
\end{tabular}
\end{table}
 
However, DL systems can be particularly challenging in that they
have variables that use sources of randomness to improve the performance of the model~\cite{gallicchio2017randomized}. In some cases, these variables are easy to identify and set (e.g. random weights initialization, batch size), in others they are easy to identify but difficult to anticipate their impact (e.g., data shuffling, dropout), and in other cases they are not even easily identifiable (e.g. more obscure options of core libraries).\footnote{For a detailed analysis   see~\cite{hung2020problems,summers2021nondeterminism,zhuang2022randomness}.} Traditionally, the ML community has focused on   classical notions of variance  associated to the dataset variables, mostly ignoring the other types~\cite{bouthillier2021accounting}.
We now analyze whether such trend also applied to the DNN4SE experiments we analyzed. Since all experiments we have studied neither explicitly analyze the sources of randomness present in the experiment, discussing how they have incorporated them into the design, nor provide the experimental design and its rationale in the paper,  the results presented here are deduced from the papers.

Our findings confirm that most experiments (62\%) acknowledge the \textbf{classical} ML random variables related to the dataset. For example papers~\citeAP{paper167}, \citeAP{paper60}, and \citeAP{paper58} use several test sets. While paper~\citeAP{paper181} uses k-fold-cross-validation. However, this leaves free other sources of randomness. We also find that some  experiments (38\%) neglect to mention \textbf{any kind of source of randomness}, approaching their experimental design with the assumption that all variables can be held constant. All these papers train the DL algorithm once, measuring the response variable(s) for the test set. An example is~\citeAP{paper09} mentioned in the previous section and also~\citeAP{paper21} which proposes a DNN that automatically applies code changes implemented by developers during pull requests (PRs). None of the papers we analyzed deal with \textbf{random variables extrinsic to the dataset}. One example of this deficiency is how  all experiments train the DNN only once for a given combination of hyperparameters. For example, In the optimization experiment reported in~\citeAP{paper30}, Xavier initialization of parameters is used. However, since the DNN is trained just once, it is impossible to know if the best configuration is due to the combination of levels of factors or just a fortuitous (random) selection of initial weights.

It is important to note that all previous instances deal with \textit{known} sources of randomness (cases II-V from Table~\ref{tab:dealing_extraneous}). But there might be \textit{unknown} sources of randomness in an experiment (case I). The ML community has not fully acknowledged the  existence of these variables, but it would be valuable for the experiments designs to safeguard against them. These variables are typically addressed by randomly assigning  the order in which the experimental runs will take place. Imagine a situation where caching is in effect for the non-initial runs. If the runs are not  randomly executed, and there is not enough of them, the first runs could behave differently from the rest. If we are comparing 2 DNNs and we plan all the runs for one of them first, this could be affecting the results. 

\subsection{Does the (Cause) Operationalization Accurately Represent its Construct? (Construct Validity)}
\label{sec:construct}

A construct validity is an assessment of how well researchers  translate their ideas  into specific factors and treatments, and response variables~\cite{wohlin2012experimentation}. Since the experiments we have analyzed operationalize well their response variables (76\% fully address it), we will focus on factors and treatments. 

The positive news is that only 4\% of experiments have a \textbf{definition of factors that is incomplete}. This is the case for many hyperparameter optimization experiments, which are often not fully acknowledged  in the papers. For example, in~\citeAP{paper05}, the hyperparameters fine-tuning optimization experiment is mostly absent. The paper briefly mentions the range of hyperparameters, and gives some examples, but the listing is not exhaustive so in the end it is not known what factors were explored. 

On the negative side, 82\% of experiments define their factors properly, but their \textbf{treatment definitions are incomplete}. For example, the optimization experiment in~\citeAP{paper76} mentions that the hyperparemeters to be fine-tuned are embedding size, number of hidden states, batch size, maximum number of iterations, optimizer, learning rate, beam size and lambda. However, it does not specify the range of values that have been explored. In~\citeAP{paper26}, the regularization term, the number of iterations, or the topology of the proposed DNN are not reported. In the experiment in~\citeAP{paper59}, the treatments are defined at the architectural level (a deep adaptation network is compared against a deep belief network, a LSTM, and a CNN). However, specific implementations of these architectures are being compared, overlooking the fact that other non-specified variables like the model hyperparameters, the DL algorithm or the data representation could be the underlying causes for the performance gain, and not the architecture.

This issue gets magnified as the limitations propagate across papers. For example,  ~\citeAP{paper05}  has an ambitious agenda to  compare the proposed technique against three other state-of-the-art approaches, but none of them are easily and reliably available. For one of them, the stable version in Github is available, but it may be different from the one in the paper (according to our results from Section~\ref{sec:analysis_artifacts}). For another, the authors of the paper had to resort to reimplement the approach following the original paper where it is proposed, which may implement a different technique. For the third one, the performance numbers reported in the original paper are used, but even if the training has ben imitated, those may have suffered from extraneous variables that are unstated. The limited availability of high-quality artifacts remains an ongoing challenge. 

Finally, 14\% of experiments had all \textbf{factors and treatments  fully operationalized}. For example, paper~\citeAP{paper60} specifies that the factors are: Word2vec vector length (with values 100, 50, 120), learning rate (with values 0.001, 0.005, 0.01) and epoch size  (with values 100, 200, 300).

\subsection{Characterization of Experiments and Implications}

\begin{figure}[bt]
      \centering
      \includegraphics[width=0.9\linewidth]{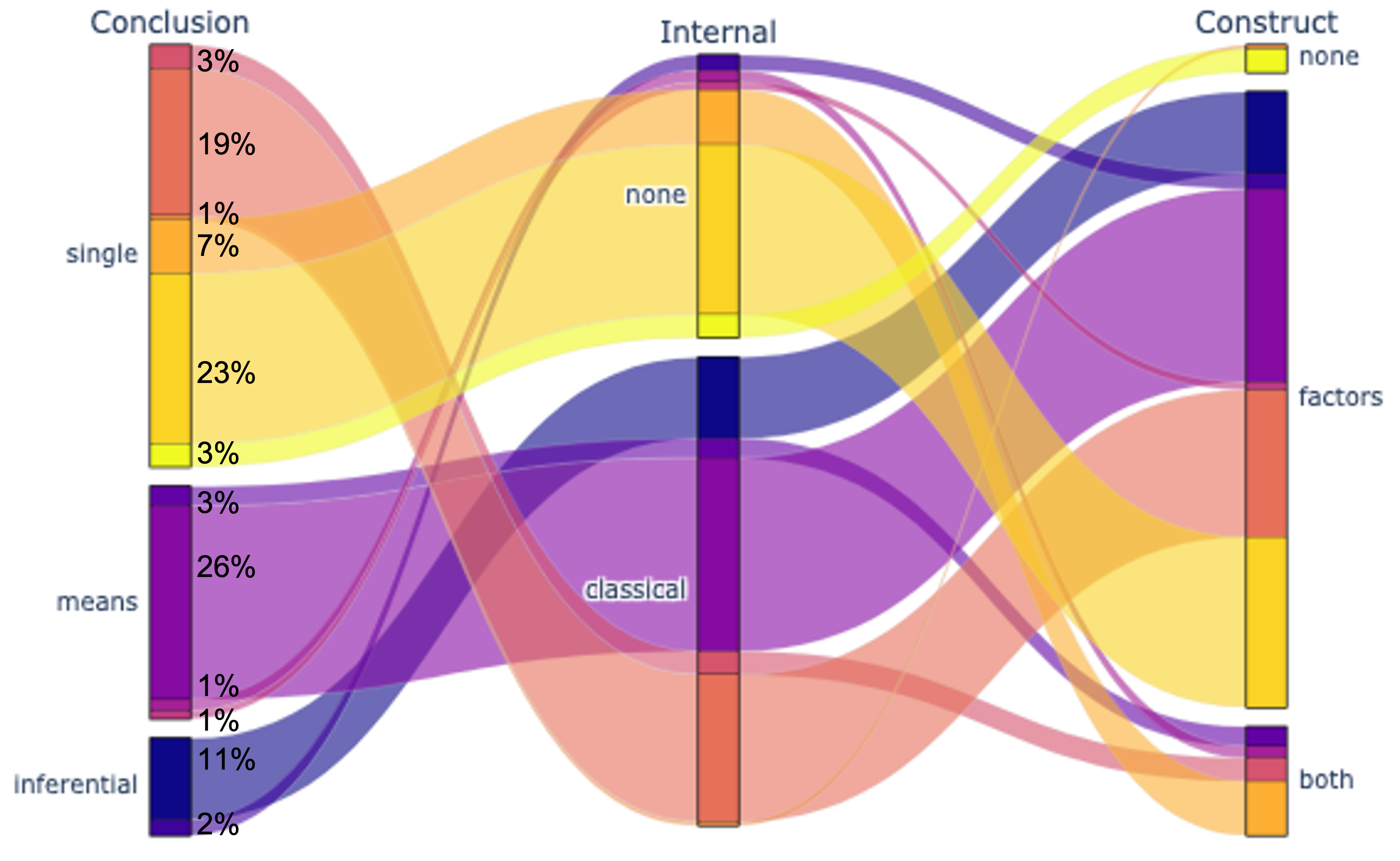}
      \caption{Distribution of experiments' implications.}
      \label{fig:bubble} 	
    \end{figure}

We now proceed to analyze the distribution of experiments'   implications to better understand how often they occur and what combinations are the most common. We utilize the parallel categories diagram in Figure~\ref{fig:bubble} to facilitate the exposition.
The sets of nodes being considered are associated with the three  implication types analyzed in the previous sections. That is, for relationship exploration (conclusion validity) we consider comparisons among single values, means, and inferential; for causality (internal validity) we consider when none, classical, and other sources of randomness are controlled; for construct validity we consider when none, factors, or both treatments and factors are defined. In the figure, the nodes on the left correspond to comparisons, the ones on the center to causality, and the ones on the right to constructs issues. 

We have not found any experiment that properly addresses all types of validity threats discussed under the implications. The best conducted experiments,  11\% of the ones examined, perform inferential analysis, control classical sources of randomness, and specify factors.  The  majority of experiments (61\%), however, have at least one critical issue (either compare single values, do not control any single source of randomness, or specify neither factors nor treatments). Even though most experiments specify at least factors, they perform comparisons based on single values and/or do not control any variables (45\%).


\section{Validity Threats}
\label{sec:validity}

We briefly discuss the main limitations arising from the scope, design, and implementation of our study~\cite{kitchenham2023segress}.

The \textbf{external validity} of our study is determined by the \textit{eligibility criteria} we chose. We concentrated on the `top' conferences and journals (Section~\ref{sec:scope}) with the expectation that the findings would constitute an upper bound for the average quality of experiments appearing in other venues.
The time period covered (2018--2021) allowed us to determine the status of recent research in the topic (2022 papers were not examined as the search was performed in 1/2022, and the analysis was conducted during 2022). Given the number of relevant papers in 2021, we randomly selected a subset to be examined. The same quality-driven and cost-control reasoning applies for us to target the artifacts with an ACM badge (Section~\ref{sec:scope_artifacts}).  

\textbf{Internal validity}. Our \textit{source of information} (Section~\ref{sec:scope}) to identify the papers---SCOPUS---included the chosen venues in the time period covered; this reduced the possibility of omitting potentially relevant studies. The \textit{search strategy} we followed is also repeatable. 
The paper \textit{selection process} required for each paper to be examined by one of the two authors; however, joint checks and discussion of papers that did not fit in existing filtering criteria reduced the chances of missing potentially relevant papers.

Due to the high data extraction costs from papers, the \textit{data collection process} (Section~\ref{sec:analysis_process} and Section~\ref{sec:analysis_process_artifacts}) considered all 17 ICSE papers 2018--2020, which were jointly examined by both authors to ensure that the collection strategies and results were aligned, while the remaining papers were examined by just one author. Again, joint checks and discussions of studies that did not fit the schema, introduced new ML elements, or had ambiguous specifications reduced possible researcher bias. Finally, doing a \textit{critical appraisal of individual sources of evidence}, we note that analyzing papers was challenging given the diversity of presentation styles, the number and complexity of the variables to check, and the increasing richness of the DNN domain. Furthermore, analyzing artifacts was a consistently arduous re-engineering process. The nature and magnitude of these analyses may have introduced errors in our measures. We attempted to control these internal threats by sharing all the intermediate results of the study with the community.

\textbf{Construct validity}. The characterization schema (\textit{data items} defined in Section~\ref{sec:analysis_process}) was specifically developed for this research. We created it starting from the steps of the experimental process and the aspects of the experiments that have to be covered during each step. Beginning with the generic definitions given by the experimental software engineering literature, the authors iteratively and systematically tailored it to the DL domain. We believe this provides a reasonable operationalization, one that is transparent as well for others to assess, refine, and reuse. A simpler assessment would just analyze the validity threats reported by the papers. However, the description of threats is typically ad-hoc and often incomplete~\cite{sjoberg22construct,ampatzoglou19identifying}. For this reason, we decided to assess the validity of the experiments from their description (and code artifacts in some cases). The \textit{syntheses of results} made in Section~\ref{sec:findings_papers} and Section~\ref{sec:findings_artifacts} allow identifying the validity level of the results reported in the studies.  


\section{Recommendations}
\label{sec:recommendations}
 
Failing to address the limitations we identified in the state of the practice could undermine much of the research devoted to DNN4SE. Thus, we propose three actionable recommendations that have the potential to address most of the pressing concerns we discovered.  

\textbf{Rec\#1: Perform Multiple DNN Training Runs to Control for Randomness.} Experiments must  strive to control  the randomness of the DNN training process. This process can introduce various sources of randomness, and a fundamental one is the random data selection and shuffling that occurs iteratively to compute the gradient over the DNN, which means that the resulting values may change over different runs. Yet, none of the papers reported to make multiple training runs to control for this  intrinsic  source of DNN randomness. This  raises questions about whether most results are caused by just a fortuitous or unfortunate   sample selection while searching for the gradient. There are other sources of randomness to consider (e.g., the initialization weights, the data splitting) but based on our findings we argue that simply conducting multiple runs of the DNN training process would enable the control of a sizable portion of the  randomness we observed. Furthermore, given the size of the experiments we analyzed and the magnitude of free  computing resources available, we found no compelling argument for not running an experiment  multiple times to account for the randomness in the DNN training process. This recommendation could benefit almost all experiments we analyzed.

\textbf{Rec\#2: Compute Statistics Over Multiple Runs and Data Partitions.} Experiments are meant to establish relationships between factors and response variables. Our analysis, however, found that 56\% of experiments identified a relationship based on single observations. Some of those studies had multiple observations gathered over multiple units of analysis (i.e., projects, releases, apps); in such cases it is difficult to justify why select a single data point to compare treatments. For the rest of the cases, however, there are plenty of opportunities to collect multiple observations. For instance, recommendation Rec\#1 for conducting multiple training runs will enable the collection of multiple observations. A second easily accessible source of observations for most of the papers we analyzed are the multiple partitions of the dataset used as part of the training. Given the number of sources of randomness in DNN training rendering multiple observations,  we find no compelling reason not to require at least a comparison of means from such observations, and if there are enough observations computer inferential statistics to judge whether the results generalize from the sample to the population. This recommendations could  benefit from 56\% to 87\% of the experiments we analyzed. 

\textbf{Rec\#3: Specify DNN Training Parameters Treatment Space  and Check for Paper Consistency Against Artifact.} We have already described the large DNN configuration space and how  different instantiations of it can dramatically impact the performance of  DNN4SE techniques. Yet, most papers fail to provide a specification of even some of the basic DNN parameters in that space. That lack can be mitigated by artifacts with code implementing the DNN training process. However, re-engineering the experimental design from such artifacts puts an undue load on the reader  and it is fault prone (we have done 44 of them to attest to that!). Furthermore, some experiments are often missing in the artifacts and it is common to find  inconsistencies between papers and their corresponding artifacts. We recommend that papers shall provide a tabular description of the DNN configuration space explored for each experiment (as we have done for each experiment analyzed -- see Appendix for samples). We also recommend for the adoption of ML experiment management tools (e.g., jupiter, mlflow, DVC) to track the DNN experiments, how they evolve, and also to control how they are shared in the papers and in the artifacts to facilitate the detection of inconsistencies. This recommendation could  benefit 86\% of the analyzed experiments.

\textbf{Deploying Mediums.} The previous recommendations can be implemented through different mediums. They can go directly to authors as part of a call for papers checklists~\cite{AAAI2023checklist,ML2020checklist,NeurIPS2022checklist}, be integrated as a part of the artifact verification process,  be provided to reviewers to help them judge a paper soundness and verifiability,  become part of broader guidelines such as the recently introduced empirical processes guidelines~\cite{ralph2020guidelines}, or serve as instructions for newcomers to the area. Given the increasing number of DNN4SE papers (the trend from Table~\ref{tab:experiments_found} indicates that they are likely to become a dominant research thrust in the venues we studied for years to come) and the pitfalls we observed and quantified, pursuing several of these mediums seems warranted. 

 \textbf{Periodic Checks of DNN4SE Paper Experiments.} Quantifications and reflections of where we stand as a community, like we have completed here, are an essential measurement stick to judge progress. Given the issues we found and the nature of DNN4SE that includes rapidly evolving technology, researchers, and methodologies, follow up checks seem required to at least determine the trends over the concerns. To reduce the cost of such checks, the framework we have defined in our evaluation could be reused and a smaller sample of the yearly experiments could be analyzed. 
\section{Related Work}
\label{sec:related_work}

A series of studies have analyzed the \textbf{quality of SE experiments}. Table~\ref{tab:rw} shows the  number of papers examined, the period covered, whether all or just a subset of papers are examined, the population the papers belong to (selected journals and/or conferences, particular conferences, or all), the sampling performed (Exhaustive or Random), the type of experiments included (human-oriented, technology-oriented, or both), and the quality aspect under examination for each of those studies. These studies differ primarily from ours in  that they focus mostly on a single quality aspect at a high-level of abstraction that is common across multiple software engineering domains, while we performed  a deeper specialized analysis on more quality aspects but focused on a single domain. 

\begin{table}[bt]
{ 
  \caption{Studies analyzing the quality of SE experiments. 
  }
  \label{tab:rw}
    \small
\setlength{\tabcolsep}{2pt}
  \begin{tabular}{ccccccl}
    \toprule

Study & Size & Period  & Population & Sample & Type & Aspect \\ \midrule

~\cite{dyba2006power} & 103 & 93--02 & Selected Js\&Cs &  E & Both & Statistical power\\
~\cite{kampenes2007effectsize} & 103 & 93--02 & Selected Js\&Cs  & E & Both & Effect size\\
~\cite{hannay2007theory} & 103 & 93--02 & Selected Js\&Cs &  E & Both & Theory\\

~\cite{jorgensen2016incorrect} & 150 & 02--12 & All &   R & Both & Researcher and \\

& & & & & & publication bias \\

~\cite{reyes2018statistical} & 51 & 06--15  & ICSE & R & Both & Correctness of \\
& & & & & & analysis \\

~\cite{sjoberg22construct} & 83 & 15--19  & Selected Js &   E & Human  & Construct validity \\ 

    \bottomrule
    \end{tabular}
} 
\end{table}

To improve the quality of experiments, the SE community has developed an extensive body of knowledge, some of which has resulted in \textbf{guidelines} for running and reporting experiments. Some of the guidelines are \textbf{general} enough to apply to any SE experiment~\cite{juristo2013basics,wohlin2012experimentation,jedlitschka2008reporting}, and therefore served as a starting point to characterize our experiments (Section~\ref{sec:analysis_papers}). 
However, such general guidelines do not address the specific challenges associated with experiments in the DNN4SE domain which is rapidly evolving and acquiring a critical momentum in the SE community. Other guidelines are \textbf{specific}. For example, there are domain-specific guidelines for the analysis of randomized testing algorithms~\cite{arcuri2011practical}, for addressing the diversity of the projects from which to get the dataset to be used in MSR studies~\cite{nagappan2013diversity}, and there are guidelines that are specific to conducting human-based experiments~\cite[Experiments]{ralph2020guidelines} or to perform benchmarking~\cite[Benchmarking]{ralph2020guidelines}. Again, although helpful, they are not addressing specific concerns raised when conducting experiments in the DNN domain. This is the first paper that characterizes the state of the practice in DNN4SE experimentation.

\section{Conclusions}
\label{sec:conclusions}

The SE community is increasingly developing techniques based on DNNs to solve software engineering problems. Performing experiments to assess such techniques is challenging given DNNs' inherent complexity involving many subtle and interdependent training variables, sources of randomness, and rapid technological evolution. Our examination of 194 experiments in 55 papers is the first to quantify these challenges. We find that 87\% of experiments are missing inferential statistics and 56\% are missing even basic descriptive statistics, 4\% are not stating the experimental factors and 82\% only do so partially, and 38\% do not specify even the basic elements of the experimental design to control any source of randomness 
while the rest only control for the classical sources of randomness. These findings' trends only mildly change   when artifacts are provided as part of such experiments, and what is more concerning is that we find that most artifacts are not fully consistent with their corresponding paper.

These findings are problematic because they imply that: 1) there is weak support to determine that there is a true relationship between independent and dependent variables that did not take place by happenstance, 2) there is limited control over the space of DNN relevant variables, which can render a relationship between dependent variables and treatments that may not be causal but rather correlational, and 3) there is a lack of specificity in terms of what are the DNN variables and their values utilized in the experiments to define the treatments being applied, which makes it unclear whether the techniques designed are the ones being assessed. We have proposed a series of actionable recommendations addressing the most critical findings we uncovered and will push forward to have them become a part of  our community practices.

\section{Data Availability}
\label{sec:repo}

The data of our analyses is currently in an anonymous repository.

\section{Acknowledgements}
\label{sec:ack}

This work was supported in part by NSF grant \#1900676. The authors are thankful to the reviewers for their feedback and to the authors of the papers and artifacts we analyzed.



\bibliographystyle{ACM-Reference-Format}
\bibliography{biblio}

\bibliographystyleAP{ACM-Reference-Format}
\bibliographyAP{analyzed_papers}


\end{document}